\documentclass[usenames,dvipsnames]{article}
\usepackage{color}

\def\be{\begin{eqnarray}}
\def\ee{\end{eqnarray}}
\def\nn{\nonumber}

\def\Tr{{\rm Tr}\,}

\definecolor{red}{rgb}{1,0,0}
\definecolor{orange}{rgb}{1,0.5,0}
\definecolor{violet}{rgb}{0.7,0,1}

\def\Color{coloring}
\def\cre{\color{red}}
\def\co{\color{orange}}
\def\cy{\color{yellow}}
\def\cg{\color{green}}
\def\cb{\color{blue}}
\def\cv{\color{violet}}

\def\trABCD{
\begin{picture}(45,20)(0,0)
\put(0,0){\mbox{$\Tr AB\bar C\bar D$}}
\qbezier(20,10)(25,15)(30,10)
\qbezier(25,-3)(30,-8)(35,-3)
\end{picture}
}
\def\trDCBA{
\begin{picture}(45,20)(0,-0)
\put(0,0){\mbox{$\Tr D C \bar B\bar A$}}
\qbezier(20,10)(25,15)(30,10)
\qbezier(25,-3)(31,-9)(37,-3)
\end{picture}
}
\def\trACBD{
\begin{picture}(45,20)(0,0)
\put(0,0){\mbox{$\Tr A\bar CB\bar D$}}
\qbezier(20,10)(22.5,13)(25,10)
\qbezier(32,-3)(35,-6)(38,-3)
\end{picture}
}

\hoffset= - 1.0in         

\begin{document}

\title{\vspace{.1cm}{\LARGE {\bf Rainbow tensor model with enhanced symmetry and extreme melonic dominance
}\vspace{.5cm}}
\author{{\bf H. Itoyama$^{a,b}$},
{\bf A. Mironov$^{c,d,e}$},
\ {\bf A. Morozov$^{d}$}
}
\date{ }
}

\maketitle

\vspace{-6.2cm}

\begin{center}
\hfill FIAN/TD-05/17\\
\hfill IITP/TH-04/17\\
\hfill ITEP/TH-09/17\\
\hfill OCU-PHYS 458
\end{center}

\vspace{4.cm}

\begin{center}
$^a$ {\small {\it Department of Mathematics and Physics, Graduate School of Science,
Osaka City University, 3-3-138, Sugimoto, Sumiyoshi-ku, Osaka, 558-8585, Japan}}\\
$^b$ {\small {\it Osaka City University Advanced Mathematical Institute (OCAMI), 3-3-138, Sugimoto, Sumiyoshi-ku, Osaka, 558-8585, Japan}}\\
$^c$ {\small {\it I.E.Tamm Theory Department, Lebedev Physics Institute, Leninsky prospect, 53, Moscow 119991, Russia}}\\
$^d$ {\small {\it ITEP, B. Cheremushkinskaya, 25, Moscow, 117259, Russia }}\\
$^e$ {\small {\it Institute for Information Transmission Problems,  Bolshoy Karetny per. 19, build.1, Moscow 127051 Russia}}
\end{center}

\vspace{.5cm}

\begin{abstract}
We introduce and briefly analyze the rainbow tensor model
where all planar diagrams are melonic.
This leads to considerable simplification of the large $N$ limit
as compared to that of the matrix model:
in particular, what are dressed in this limit are propagators only,
which leads to an oversimplified closed set of Schwinger-Dyson
equations for multi-point correlators.
We briefly touch upon the Ward identities, the substitute of
the spectral curve and the AMM/EO topological recursion
and their possible connections to Connes-Kreimer theory and
forest formulas.
\end{abstract}

\bigskip

\bigskip

\paragraph{Introduction.}
Tensor models are thought of as straightforward generalizations of matrix
models from matrices to tensors \cite{tensor}, and are the most natural objects to consider in the
study of non-linear algebra \cite{NLA}.
For a variety of reasons, they did not long attract the attention they deserve,
though a lot of important work was performed by numerous research groups (see \cite{GrFieldTh,Akhm}
and especially \cite{BGRfirst}-\cite{BGRlast} for
incomplete set of references).

The current surge of interest to tensor models \cite{Witten,GuraupostWit,KleTar,Gr}
is partly explained by their {\it simplified}
large $N$ behavior as compared to the matrix models:
instead of {\it all} planar diagrams, what contribute
in some cases are only {\it melonic} ones.
This makes the Schwinger-Dyson equation (SDE) in the large $N$ limit
almost as simple as in the leading logarithmic approximation in QED:
it is enough to dress the propagator, while the vertices
remain undressed.
The only complication is that, in this limit, the SDE is no longer
linear and solved by a geometric progression:
instead it turns into a higher order equation,
but the semi-infinite Bogoliubov chain does not arise.

In this letter, we consider the {\it rainbow} tensor
model which possesses these properties {\it in extreme}:
it selects the melonic diagrams out of all planar {\it by symmetry},
all others simply do {\it not} exist as its Feynman diagrams.
In exchange, it clarifies to some extent what is the price to pay
for this "simplification":
already the SDE becomes much more involved when one considers
a generic point in the moduli space of external parameters ({\Color}s).
The model can be easily formulated for tensors of any rank $D$,
but all its features are already seen at the level of $D=3$,
which we use in most considerations to simplify the presentation.
We mostly repeat in slightly different words and with slightly
different accents the elementary basic observations,
which were already made in the above cited references,
and try to minimize deviations from the previously introduced
notation.
We concentrate on the matrix model aspects of the story
and do not consider time-dependence \cite{Witten,GuraupostWit,KleTar,Gr},
relation to SYK model \cite{SY,K}, holography \cite{K}.

We also do not dwell upon intriguing similarity to arborescent calculus in knot theory \cite{arbor}. It is, however, reflected in some of the terminology below.

\paragraph{The tetrahedron/starfish model} is
\be
Z =   \prod_{I=0}^{D} \int d^2A_I e^{-M_IA_I\bar A_I}
\exp\left( g \prod_{I=0}^D A_I + g \prod_{I=0}^D \bar A_I\right)
\ee
Each $A_I$ is a rank $r$ tensor, and $\bar A_I$ is obtained by exchanging
covariant and contravariant indices. In $A_I\bar A_I$, the indices
are convoluted in an obvious way, while the interaction term
is a {\it "tetrahedron"} or, generically, a {\it "starfish"} vertex, with indices contracted in
a very special way:

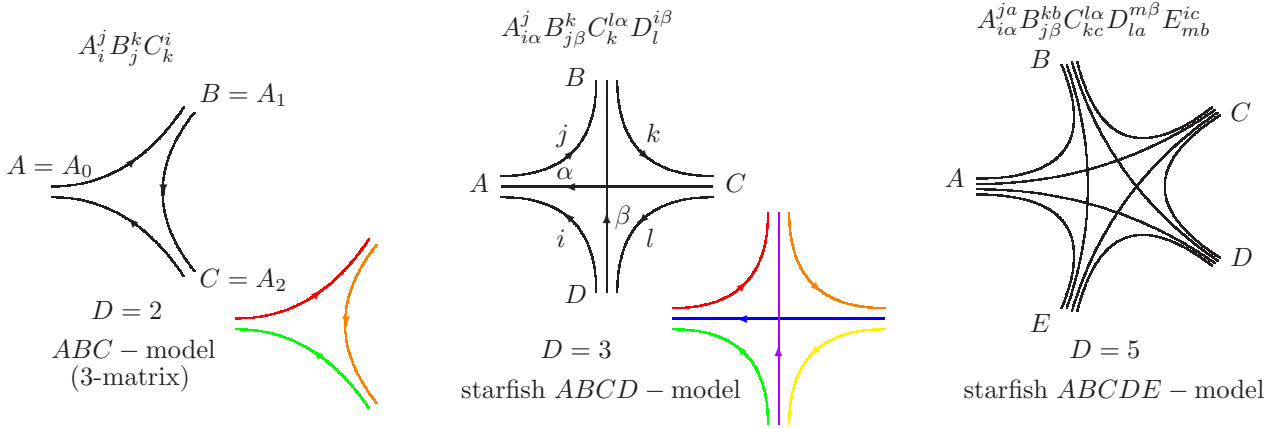
\begin{picture}(300,200)(0,-125)
\qbezier(0,0)(30,0)(50,30)\put(30,9){\vector(1,1){2}}
\qbezier(0,-4)(30,-4)(50,-34) \put(42.4,-1){\vector(0,-1){2}}
\qbezier(54,28)(30,-2)(54,-32) \put(31,-13.8){\vector(-1,1){2}}
\put(70,-50){
\put(0,0){{\color{red}\qbezier(0,0)(30,0)(50,30) \put(30,9){\vector(1,1){2}}} }
{\color{green}\qbezier(0,-4)(30,-4)(50,-34) \put(31,-13.8){\vector(-1,1){2}}}
\put(-1,0){{\co\qbezier(54,28)(30,-2)(54,-32) \put(42.4,-1){\vector(0,-1){2}} } }
}
\put(-17,5){\mbox{$A=A_0$}}
\put(56,32){\mbox{$B=A_1$}}
\put(56,-38){\mbox{$C=A_2$}}
\put(10,50){\mbox{$A_{i}^j B_{j}^k C_k^i$}}
\put(15,-50){\mbox{$D=2$}}
\put(0,-65){\mbox{$ABC-{\rm model}$}}
\put(7,-75){\mbox{(3-matrix)}}
\put(50,0){
\qbezier(120,4)(156,4)(156,40)\put(146,12){\vector(1,1){2}}
\qbezier(120,-4)(156,-4)(156,-40)\put(146,-12){\vector(-1,1){2}}
\qbezier(200,4)(164,4)(164,40)\put(174,12){\vector(1,-1){2}}
\qbezier(200,-4)(164,-4)(164,-40)\put(174,-12){\vector(-1,-1){2}}
\put(120,0){\line(1,0){80}}\put(146,0){\vector(-1,0){2}}
\put(160,40){\line(0,-1){80}}\put(160,-12){\vector(0,1){2}}
\put(65,-50){
\put(0,0){{\cre\qbezier(120,4)(156,4)(156,40)\put(146,12){\vector(1,1){2}}}}
{\cg\qbezier(120,-4)(156,-4)(156,-40)\put(146,-12){\vector(-1,1){2}}}
\put(0,0){{\co\qbezier(200,4)(164,4)(164,40)\put(174,12){\vector(1,-1){2}}}}
{\cy\qbezier(200,-4)(164,-4)(164,-40)\put(174,-12){\vector(-1,-1){2}}}
\put(0,0){\cb\put(120,0){\line(1,0){80}}\put(146,0){\vector(-1,0){2}}}
\put(0,0){\cv\put(160,40){\line(0,-1){80}}\put(160,-12){\vector(0,1){2}}}
}
\put(108,-2){\mbox{$A$}}
\put(144,38){\mbox{$B$}}
\put(205,-2){\mbox{$C$}}
\put(144,-44){\mbox{$D$}}
\put(120,57){\mbox{$A_{i\alpha}^j B_{j\beta}^k C_k^{l\alpha} D_l^{i\beta}$}}
\put(135,-65){\mbox{$D=3$}}
\put(105,-80){\mbox{starfish $ABCD-{\rm model}$}}
\put(141,17){\mbox{$j$}} \put(175,17){\mbox{$k$}}
\put(141,-23){\mbox{$i$}} \put(175,-23){\mbox{$l$}}
\put(141,3){\mbox{$\alpha$}} \put(163,-14){\mbox{$\beta$}}
}
\put(400,0){
%

\qbezier(-50,3)(0,3)(-18,46)
\qbezier(-12,47)(0,0)(39,30)
\qbezier(42,27)(0,0)(42,-27)
\qbezier(-12,-47)(0,0)(39,-30)
\qbezier(-50,-3)(0,-3)(-18,-46)
\qbezier(-50,1)(0,1)(40,29)
\qbezier(-50,-1)(0,-1)(40,-29)
\qbezier(-16,46)(0,0)(-16,-46)
\qbezier(-14,47)(0,0)(41,-28)
\qbezier(-14,-47)(0,0)(41,28)
\put(-50,60){\mbox{$A_{i\alpha}^{ja} B_{j\beta}^{kb} C_{kc}^{l\alpha} D_{la}^{m\beta} E_{mb}^{ic}$}}
\put(-15,-65){\mbox{$D=5$}}
\put(-55,-80){\mbox{starfish $ABCDE-{\rm model}$}}
\put(-62,0){\mbox{$A$}}
\put(-30,45){\mbox{$B$}}
\put(46,25){\mbox{$C$}}
\put(46,-30){\mbox{$D$}}
\put(-30,-55){\mbox{$E$}}
 %


%
}
\put(0,-100){\mbox{
\footnotesize{In general, the indices here belong to different groups
(tensors are "rectangular"): \
 $i=1,\ldots,N_{{\rm green}}$, $j=1,\ldots,N_{{\rm red}}$,}}}
\put(0,-110){\mbox{
\footnotesize{
 $k=1,\ldots,N_{{\rm orange}}$, $l=1,\ldots,N_{{\rm yellow}}$, $\alpha=1,\ldots, N_{{\rm blue}}$,
$\beta=1,\ldots,N_{{\rm violet}}$, $a=1,\ldots, N_{{\rm brown}}$, $b=1,\ldots, N_{{\rm pink}}$,
$\ldots$
}}}

\end{picture}

\noindent
When one draws a vertex for $D=3$ as contraction of four rank $D$ tensors,
it looks like a tetrahedron:

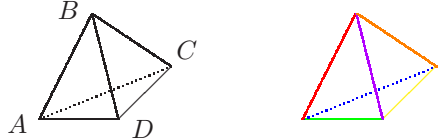
\begin{picture}(300,55)(-150,-5)
\put(0,0){\line(1,0){30}}
\put(30,0){\line(1,1){20}}
\qbezier[30](0,0)(25,10)(50,20)
\qbezier(0,0)(10,20)(20,40)
\qbezier(30,0)(25,20)(20,40)
\qbezier(50,20)(35,30)(20,40)
\put(-12,-5){\mbox{$A$}}
\put(35,-7){\mbox{$D$}}
\put(7,38){\mbox{$B$}}
\put(52,23){\mbox{$C$}}
\put(100,0){
{\put(0,0){\color{green}\line(1,0){30}}}
{\put(30,0){\color{yellow}\line(1,1){20}}}
{\cb\qbezier[30](-3,0)(22,10)(47,20)}
{\cre\qbezier(-7,0)(3,20)(13,40)}
{\cv\qbezier(20,0)(15,20)(10,40)}
{\co\qbezier(37,20)(22,30)(7,40)}
}
\end{picture}

\noindent
but if one inserts it as a vertex into Feynman diagrams, the lines at the corners
do not merge and the picture looks more like a starfish as above, which explains the two names.

In the simplest model of this type \cite{uncl},
one makes no difference between
the $D$ fields ($A=B=C=\ldots)$ and does not distinguish the upper and lower indices.
This model possesses the $O(N)^D$ symmetry with {\it limited} \Color, as we have
$D$ instead of $\frac{D(D+1)}{2}$.
It was studied in \cite{KleTar} under the name of "uncolored", though in our context  "$D$-colored" or "single-field" seems somewhat more appropriate.
Still, in the large $N$ limit, it also represents the
universality class of  models which are {\it dominated} by the melonic diagrams.

\paragraph{The  rainbow model.}
An opposite option is to provide each line in these pictures
with its own {\Color}, i.e. to endow the model with the symmetry
\be
{\cal U}_D = \prod_{a=1}^{\frac{D(D+1)}{2}} U(N_a)
\ \stackrel{N_a=N}{\longrightarrow} \ U(N)^{^{\otimes \frac{D(D+1)}{2}}}
\ee
For $D=3$, this gives six different {\Color}s, thus the name {\it rainbow},
and we sometimes denote these {\Color}s by $r$ (red), $o$ (orange), $y$ (yellow),
$g$ (green), $b$ (blue) and (missing indigo) $v$ (violet).
Accordingly, the vertices have three ($D$) {\Color}s, but not every triple is allowed:
there are just four ($D+1$) permitted combinations
\be
I \in
\{A=\bar g br,\ B=\bar rvo, \ \bar C = \bar o\bar by, \ \bar D = \bar y\bar v g\}
= \{A=\bar {\cg g}{\cb b}{\cre r},\ B=\bar {\cre r}{\cv v}{\co o},
\ \bar C = {\co \bar o}\bar {\cb b}{\cy y}, \ \bar D = \bar {\cy y}\bar {\cv v} {\cg g}\}
\ee
Each field transforms under the action of the corresponding three ($D$) groups:
\be
A_{{\color{red}i_r}\  {\color{blue}i_b}}^{\ {\color{green}i_g}}  \longrightarrow
\sum_{j_r=1}^{N_r} \sum_{j_g=1}^{N_g} \sum_{j_b=1}^{N_b}
{\cre U_{i_r}^{j_r}} {\cg \bar U^{i_g}_{j_g}} {\cb U_{i_b}^{j_b}}\
A_{{\cre j_r}\  {\cb j_b}}^{\ {\cg j_g}} , \nn \\
B_{{\co i_o}\  {\cv i_v}}^{\ {\cre i_r}}  \longrightarrow
\sum_{j_r=1}^{N_r} \sum_{j_g=1}^{N_g} \sum_{j_b=1}^{N_b}
{\co U_{i_o}^{j_o}} {\cre \bar U^{i_r}_{j_r}} {\cv U_{i_v}^{j_v}}\
B_{{\co j_o}\  {\cv j_v}}^{\ {\cre j_r}} , \nn \\
\bar C_{{\cy i_y} }^{\, {\co i_o}\,{\cb i_b}}  \longrightarrow
\sum_{j_r=1}^{N_r} \sum_{j_g=1}^{N_g} \sum_{j_b=1}^{N_b}
{\cy U_{i_y}^{j_y}} {\co \bar U^{i_o}_{j_o}} {\cb \bar U^{i_b}_{j_b}}\
\bar C_{{\cy j_y} }^{\, {\co j_o}\,{\cb j_b}}, \nn \\
\bar D_{{\cg i_g} }^{\, {\cy i_y}\,{\cv i_v}}  \longrightarrow
\sum_{j_r=1}^{N_r} \sum_{j_g=1}^{N_g} \sum_{j_b=1}^{N_b}
{\cg U_{i_g}^{j_g}} {\cy \bar U^{i_y}_{j_y}} {\cv \bar U^{i_v}_{j_v}}\
\bar D_{{\cg j_g} }^{\, {\cy j_y}\,{\cv j_v}}
\ee
It is often convenient to distinguish between the "external" and "internal" lines
in the pictures, and denote the interaction vertex by
\be
\prod_{I=0}^3 A_I
= A_{i_r\, k_b}^{i_g} B_{i_o\, k_v}^{i_r} \bar C_{i_y}^{i_o\, k_b} \bar D_{i_g}^{i_y\, k_v}
= A_{{\cre i_r}\, {\cb k_b}}^{{\cg i_g}} B_{{\co i_o}\, {\cv k_v}}^{{\cre i_r}}
\bar C_{{\cy i_y}}^{{\co i_o}\, {\cb k_b}} \bar D_{{\cg i_g}}^{{\cy i_y}\, {\cv k_v}}
= \trABCD\ ,
\ee
i.e. we consider the fields as matrices with respect to the indices $i_{r,o,y,g}$,
while the additional indices $k_b$ and $k_v$ correspond to the additional (internal) lines in the Feynman graphs.

To avoid possible confusion, we emphasize that our {\it {\Color}s} distinguish
different components of the symmetry {\it group}: they are not the {\it colors}
labeling different elements of the fundamental representation of a single $SU(N)$
(like quark colors in QCD).
One could rather associate them with a kind of {\it flavors} labeling elements of representation
of a global symmetry $O\left(\frac{D(D+1)}{2}\right)$, but the closest
is the analogy with the constituents of the gauge group in the quiver models
of \cite{Gai}.
The difference from the $D$-colored model of \cite{KleTar} is that, in the rainbow model, one has
$D+1$ different types of fields and propagators labeled by a set $I$ of indices.

\paragraph{Planar and melonic diagrams.}
This latter difference, however, has a profound effect on the calculus based on diagrams.
With the interaction vertices above and the above notation, one can formally separate the
"external" and "internal" {\Color}s and draw Feynman diagrams as the ordinary matrix model
double-line (fat graph) diagrams for the "external" ones.
In other words, the propagators are still rather bands than tubes (not all the {\Color}s are
on equal footing).
Extra internal lines are inserted at the next stage and provide an additional structure.
The point is that even if the fat graphs are planar, the loop calculus for internal lines
can still be different, and this leads to distinguishing the class of melonic diagrams from others.
Specifics of the {\it rainbow} model is that the possible contractions of internal lines
are strongly limited, so that only the melonic planar fat graphs are actually contributing, while
the non-melonic planar graphs are not just {\it damped} by powers of $N_{int}$ (as in the case of the $D$-colored model),
but they are simply absent.
The phenomenon is clear from the following picture:

\bigskip

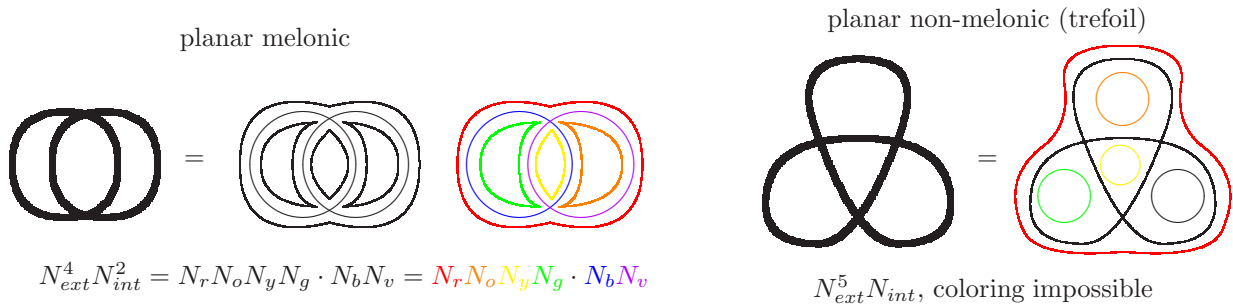
\begin{picture}(300,120)(-115,-60)
{\linethickness{0.8mm}
\qbezier(-100,-20)(-120,-20)(-120,0)\qbezier(-100,-20)(-80,-20)(-80,0)
\qbezier(-100,20)(-120,20)(-120,0)\qbezier(-100,20)(-80,20)(-80,0)
\qbezier(-85,-20)(-105,-20)(-105,0)\qbezier(-85,-20)(-65,-20)(-65,0)
\qbezier(-85,20)(-105,20)(-105,0)\qbezier(-85,20)(-65,20)(-65,0)
}
\put(-55,0){\mbox{$=$}}
\qbezier(0,22)(34,30)(34,0)\qbezier(0,-22)(34,-30)(34,0)
\qbezier(0,22)(-34,30)(-34,0)\qbezier(0,-22)(-34,-30)(-34,0)
\qbezier(5,16)(26,15)(26,0)\qbezier(5,-16)(26,-15)(26,0)
\qbezier(-5,16)(-26,15)(-26,0)\qbezier(-5,-16)(-26,-15)(-26,0)
\qbezier(5,16)(14,14)(14,0)\qbezier(5,-16)(14,-14)(14,0)
\qbezier(-5,16)(-14,14)(-14,0)\qbezier(-5,-16)(-14,-14)(-14,0)
\qbezier(0,13)(14,0)(0,-13) \qbezier(0,13)(-14,0)(0,-13)
\put(-10,0){\circle{40}}
\put(10,0){\circle{40}}
\put(80,-0){
{\cre \qbezier(0,22)(35,30)(35,0)\qbezier(0,-22)(35,-30)(35,0)
\qbezier(0,22)(-35,30)(-35,0)\qbezier(0,-22)(-35,-30)(-35,0)}
{\co \qbezier(0,16)(24,15)(24,0)\qbezier(0,-16)(24,-15)(24,0)}
{\cg \qbezier(-12,16)(-33,15)(-33,0)\qbezier(-12,-16)(-33,-15)(-33,0)}
{\co \qbezier(-6,16)(2,14)(2,0)\qbezier(-6,-16)(2,-14)(2,0)}
{\cg \qbezier(-17,16)(-25,14)(-25,0)\qbezier(-17,-16)(-25,-14)(-25,0)}
{\cy \qbezier(-16,13)(-6,0)(-16,-13) \qbezier(-16,13)(-28,0)(-16,-13)}
{\cb \put(-31.5,0){\circle{40}}}
{\cv \put(-11.5,0){\circle{40}}}
}
\put(-60,45){\mbox{ planar melonic}}
\put(-110,-45){\mbox{$N_{ext}^4 N_{int}^2
= N_rN_oN_yN_g\cdot N_bN_v
= {\cre N_r}{\co N_o}{\cy N_y}{\cg N_g}\cdot {\cb N_b}{\cv N_v}$
}}
\put(200,10){
{\linethickness{0.8mm}
\qbezier(-35,-25)(-35,0)(0,0)\qbezier(35,-25)(35,0)(0,0)
\qbezier(-35,-25)(-35,-40)(-15,-40)\qbezier(35,-25)(35,-40)(15,-40)
\qbezier(-15,-40)(0,-40)(15,0)\qbezier(15,-40)(0,-40)(-15,0)
\qbezier(15,0)(27,30)(0,30)\qbezier(-15,0)(-27,30)(0,30)
}
}
\put(245,0){\mbox{$=$}}
\put(300,0){
\put(0,10){
\qbezier(-35,-25)(-35,0)(0,0)\qbezier(35,-25)(35,0)(0,0)
\qbezier(-35,-25)(-35,-40)(-15,-40)\qbezier(35,-25)(35,-40)(15,-40)
\qbezier(-15,-40)(0,-40)(15,0)\qbezier(15,-40)(0,-40)(-15,0)
\qbezier(15,0)(27,30)(0,30)\qbezier(-15,0)(-27,30)(0,30)
\put(-0.8,0){{\cy\put(0,-10){\circle{15}}}}
{\co\put(0,15){\circle{20}}}
\put(-1,0){{\cg\put(-21,-22){\circle{20}}}}
\put(21,-22){\circle{20}}
{\cre
\qbezier(-38,-34)(-34,-46)(0,-43)
\qbezier(-38,-34)(-46,-15)(-30,-3)
\qbezier(-30,-3)(-20,5)(-22,20)
\qbezier(-22,20)(-25,35)(0,35)
\qbezier(38,-34)(34,-46)(0,-43)
\qbezier(38,-34)(46,-15)(30,-3)
\qbezier(30,-3)(20,5)(22,20)
\qbezier(22,20)(25,35)(0,35)
}

}
\put(-115,52){\mbox{ planar non-melonic (trefoil)}}
\put(-117,-50){\mbox{$N_{ext}^5 N_{int}$,\  coloring impossible}}
}
\end{picture}

\noindent
It is clear that, in the second non-melonic picture, the
internal line makes just one loop instead of three,
which causes a damping factor $N_{int}^{-2}$.
However, such an internal line can not be ascribed any
definite {\Color} (it self-intersects, thus can be neither
blue nor violet), i.e. such a diagram simply does not exist
in the {\it rainbow} model (and exists in the $D$-colored model, where it is dumped in large $N$ limit).
This is an illustration of the general statement:
while in all tetrahedron/starfish models beginning from the $D$-colored one,
the melonic diagrams are the only surviving in the large $N$ limit, while
{\bf in the rainbow model, all the planar diagrams are automatically melonic}.

Moreover, even the right internal circle in the trefoil picture
can not be ascribed any definite coloring: should it be orange or green?
However, this is
because the trefoil is not just non-melonic, but it also has an
odd number of vertices, thus it is forbidden in the rainbow model by a much simpler reason:
each  diagram should have equal number of chiral and anti-chiral vertices $\trABCD$ and $\trDCBA$.
We have, however, emphasized another argument related to the absence of
$\trACBD$ vertices with another contraction of indices, which remains applicable
to the case of non-melonic diagrams with an even number of vertices as well: the following diagram

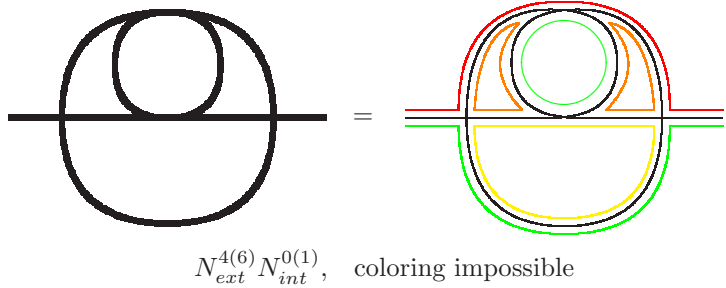
\begin{picture}(300,120)(-250,-70)
\put(-40,-2){\mbox{$=$}}
{\cre \put(0,3){\line(-1,0){20}}\qbezier(0,3)(0,44)(40,44)
\qbezier(40,44)(80,44)(80,3)\put(80,3){\line(1,0){20}}  }
\put(0,0){{\cg \put(0,-3){\line(-1,0){20}}\qbezier(0,-3)(0,-44)(40,-44)
\qbezier(40,-44)(80,-44)(80,-3)\put(80,-3){\line(1,0){20}}  }}
{\cy \put(6,-3){\line(1,0){68}} \qbezier(6,-3)(6,-38)(40,-38)\qbezier(40,-38)(74,-38)(74,-3) }
\put(-0,0){{\cg \put(40,21){\circle{32}}}}
%
{ \put(-20,0){\line(1,0){55}} \put(45,0){\line(1,0){55}} }
\qbezier(35,0)(60,0)(60,21)\qbezier(45,0)(20,0)(20,21)
\qbezier(20,21)(20,41)(45,41)\qbezier(60,21)(60,41)(35,41)
\qbezier(3,0)(3,41)(35,41)\qbezier(77,0)(77,41)(45,41)
\qbezier(3,0)(3,-41)(40,-41)\qbezier(77,0)(77,-41)(40,-41)
{\co  \put(6,3){\line(1,0){18}}\qbezier(6,3)(8,33)(23,36) \qbezier(24,3)(8,20)(23,36)
\put(74,3){\line(-1,0){18}}\qbezier(74,3)(72,33)(57,36) \qbezier(56,3)(72,20)(57,36)}
\put(-150,0){
{\linethickness{0.7mm}
\put(-20,0){\line(1,0){120}}
\qbezier(0,0)(0,40)(40,40)\qbezier(80,0)(80,40)(40,40)
\qbezier(0,0)(0,-40)(40,-40)\qbezier(80,0)(80,-40)(40,-40)
\qbezier(40,0)(20,0)(20,20)\qbezier(20,20)(20,40)(40,40)
\qbezier(40,0)(60,0)(60,20)\qbezier(60,20)(60,40)(40,40)
}
}
\put(-100,-60){\mbox{$N_{ext}^{4(6)}N_{int}^{0(1)}$, \ \ coloring impossible}}
\end{picture}

\noindent
is also an impossible picture in the rainbow model: there is no way to ascribe colors
(blue and violet) to the black internal line.
The powers in brackets correspond to the case of the vacuum diagram,
when the external lines are connected to form two additional loops, and
non-melonic damping is then due to the first (instead of the third) power of $N_{int}$.

\paragraph{SDE in the melonic limit.}
The dominance of melonic diagrams leads to a very simple Schwinger-Dyson equation
for the dressed propagator:

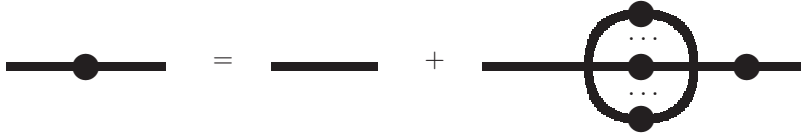
\begin{picture}(300,100)(-50,-50)
\linethickness{1mm}
\put(0,0){\line(1,0){60}}
\put(30,0){\circle*{10}}
\put(78,0){\mbox{$=$}}
\put(100,0){\line(1,0){40}}
\put(158,0){\mbox{$+$}}
\put(180,0){\line(1,0){120}}
%
\qbezier(240,-20)(220,-20)(220,0)\qbezier(240,-20)(260,-20)(260,0)
\qbezier(240,20)(220,20)(220,0)\qbezier(240,20)(260,20)(260,0)
\put(240,20){\circle*{10}}
\put(240,0){\circle*{10}}
\put(240,-20){\circle*{10}}
\put(280,0){\circle*{10}}
\put(235,10){\mbox{$\ldots$}}
\put(235,-11){\mbox{$\ldots$}}
\end{picture}

\noindent
The  lines in this picture are drawn thick to emphasize that
they are now $D$-colored, i.e. labeled by indices $I$.
However, the tensor structure of lines is simply $\delta_{IJ}$,
while the change of multi-{\Color}s in the vertices is described as above.
Simplicity of the SDE equation is due to the lack of any dressing at vertices.
If we denote the dressed propagators by $G_I$, then the SDE
in the large $N$ approximation turns into
\be
G_I = 1 + g^2\prod_{a\notin I} N_a \cdot G_I \prod_{J\neq I} G_J =
1 + \frac{g^2}{N_I}\cdot\left( \prod_{a=1}^{\frac{D(D+1)}{2}} N_a \cdot \prod_{J=1}^D G_J\right)
\ee
where $N_I = \prod_{a\in I} N_a$.
If one does not differentiate the {\Color}s, then this leads us to the large $N$
behavior of the type $G\sim N^{-\frac{(D-1)D}{2(D+1)}} \ \stackrel{D=3}{=} N^{-3/4}$,
and the term at the l.h.s does not contribute.
However, when {\Color}s are different, as in the rainbow model, things are not so simple:
the factor in the bracket at the r.h.s. is independent of $I$, while the coefficient in front,
$N_I^{-1}$ depends on $I$, thus, the solution with the l.h.s. neglected can not exist.

To see what happens in this case, one can consider the simplest formal example of $\# I = 2$
\be
G_1 = 1 - \alpha_1 G_1G_2 \nn \\
G_2 = 1 - \alpha_2 G_1G_2
\ee
where the solution is just
\be
 G_1 =1-
\frac{\alpha_1+\alpha_2+1 \pm \sqrt{(\alpha_1-\alpha_2)^2+2(\alpha_1+\alpha_2)+1}}{2\alpha_2}
\\
G_2 =1-
\frac{\alpha_1+\alpha_2+1 \pm \sqrt{(\alpha_1-\alpha_2)^2+2(\alpha_1+\alpha_2)+1}}{2\alpha_1}
\ee
Then, for the sake of definiteness, choosing the plus sign in front of the square root and $\alpha_1>\alpha_2$, we obtain
\be
 G_1 \sim 1-\frac{\alpha_1}{\alpha_2}+ O\Big(\alpha^{-1}\Big)\nn \\
G_2 \sim -\frac{1}{\alpha_1-\alpha_2}+O\Big(\alpha^{-2}\Big)
\ee
or vice versa, under $\alpha_1\longleftrightarrow\alpha_2$ or under the other sign.
We are interested in the limit of large $\alpha$, and we see that the solution
depends strongly on the ratio $\alpha_1/\alpha_2$.
For bigger values of $D(\geq 3)$, more different branches appear. At the transition lines from one to the other branch, there is a multicritical behaviour. In the example of $D=2$, it is just the critical point\footnote{In terms of the group ranks, this critical point corresponds to $N_1=N_2$. This reminds the phase transition when the number of SYK Majorana fermions matches the number of free Majorana fermions in \cite{R2}, and also the phase transition in \cite{R1} when the number of Majorana fermions matches the number of replicas. We are grateful to the reviewer of our paper who paid our attention at this fact.} $\alpha_1=\alpha_2=\alpha$ where both
\be
G_i\sim -{1\over 2\alpha}+O\Big(\alpha^{-2}\Big)
\ee
At $D=3$, there are already a few critical lines (such as $\alpha_2=\alpha_3$, $\alpha_1^2-\alpha_1\alpha_2-\alpha_1\alpha_3+\alpha_2\alpha_3$ etc). This means that changing the size of the gauge group is related to a relevant operator, and there are some marginal deformations that act along the critical surfaces. One should proceed further with studying the wall-crossing phenomenon in this case.
Thus, {\bf the large $N$ limit in the rainbow model is very singular}:
it non-analytically depends on the ratios of the {\Color}s.
On one hand, this looks very similar to singularities of conformal blocks
in the vicinity of singular points (zeroes of Kac determinant) considered in \cite{IMM}.
On another hand, this looks like a peculiar feature of the dominance of melonic diagrams,
which allows dressing of propagators only (not vertices) and leads
to the oversimplified SDE.
In large $N$ matrix (not tensor) models, which are dominated by {\it all} planar diagrams,
like $ABC$-model at $D=2$,
such singularities seem to be absent. Simple examples are the
vector models, which are of course very distinguished
among the models of rectangular matrices \cite{recMM}, but this is a limit of $N_1\ll N_2$, not $N_1\sim N_2\gg 1$.
Likewise, such singularities are most probably absent in the quiver Yang-Mills models.
In this sense, the tensor models can get much closer to conformal theories (possessing a {\it simpler} large $N$ limit)
as compared to the matrix and Yang-Mills models, which can also be a manifestation of their
close relation to holography \cite{K}.

The melonic {\it dominance} (which, in the case of rainbow model, becomes {\it absolute})
implies that {\bf dressed in the large $N$ limit are only propagators}, not vertices,
which makes this part of the theory essentially {\it Abelian}
(and similar to QED in the leading logarithmic approximation).
In particular, all multi-point correlators can be easily expressed through the
dressed propagators, as is already discussed in \cite{KleTar} (where the propagators
are not-quite-trivial given the time-dependence such as in \cite{Witten}, and this would presumably put the model in the universality class of the quenched SYK models
\cite{SY,K}.)

\paragraph{Single-trace operators.}
A very first step in the matrix-model-based study of effective actions \cite{UFN3}
is division of all observables (gauge invariant operators) into {\it three} classes.
First of all, the set of operators forms a ring, where addition and multiplication
are defined (the latter one, operator product expansion can be quite involved beyond matrix models),
and one distinguishes linearly generated
operators from multiplicatively generated ones: roughly speaking, linearly independent are all gauge invariant multi-trace
combinations $\prod_{a=1}^m Tr M^{k_a}$ with any $m$ (we denote the corresponding couplings in the generating functional of all correlators as $t_{k_1k_2\ldots k_m}$),
while multiplicatively independent ones are just single traces: $\Tr M^k$. From the point of view of effective actions, the difference between them is
that dependence on the first ones is easily reduced to that on the second:
\be
\frac{\partial Z}{\partial{t_{k_1k_2\ldots k_m}}} =
\ \left<\prod_{a=1}^m \Tr M^{k_a}\right> \ = \frac{\partial^m Z}{\partial t_{k_1}
\partial t_{k_2}\ldots \partial t_{k_m}}
\label{derder}
\ee
i.e. introduction of all the $t_{k_1k_2\ldots k_m}$ into the set of couplings is
superficial: the set $\{t_k\}$ is quite enough.
Note that this does not necessarily mean that the averages factorize as
this happens only in the large $N$ limit:
\be
\left<\prod_{a=1}^m \Tr M^{k_a}\right> \
\stackrel{{\rm at\ large} \ N}{=} \ \prod_{a=1}^m \left<  \Tr M^{k_a}\right>
\ee
Hence, there is no need to introduce multi-trace couplings in order to generate all correlators
at any $N$.

Dependence on single-trace operators is more involved.
It is controlled by {\it the Ward identities}, the quantum substitute
of classical equations of motion in matrix models, reflecting the invariance
of (functional) integral under the change of integration variables.
These are nicknamed as Virasoro/W-constraints,
because they are just these in the simplest possible examples like
\be
\left<\Tr W'(M)+kt_k \Tr M^{k+n} + \sum_{a+b=n} \Tr M^a \Tr M^b\right> = 0, \ \ \ \ n\geq -1
\label{virc}
\ee
for the Hermitian matrix model. Here $W(M)$ is the potential (action) of the model.
What makes it non-trivial is the last term, which disappears
only when there are multiplicative dependence {\it among} the "single-trace"
operators. This happens for vector models, when the rectangular matrices
$M$ degenerate to vectors, and $\Tr (M\bar M)^k = (\vec V^2)^k$.
The situation is a little more involved and interesting for rectangular
matrices of generic size \cite{recMM}.

Unfortunately, in multi-matrix models, of which the {\it rainbow tensor} model
is a generalization, the set of single-trace operators becomes
unacceptably big.
Already for the  $ABC$-model, the naive set
consists of "chiral" ones $\Tr (ABC)^k$,
"antichiral" ones $\Tr (\bar A\bar B\bar C)^k$,
non-chiral ones $\Tr (A\bar A)^k,\ \Tr (B\bar B)^k, \ \Tr(C\bar C)^k$
and their arbitrarily sophisticated mixtures such as
\be
\Tr \left\{(ABC)^{k_1}(A\bar A)^{m_1} \Big(AB(C\bar C)^{l} C\Big)^{n_1}
(A\bar A)^{m_1}(ABC)^{k_2}(A\bar A)^{n_2}\ldots\right\}
\label{complictedsingletrace}
\ee
Thus, one is led to further division of single-trace operators into two classes:
chiral and non-chiral.

\paragraph{Chirality.} The interaction vertex in the $ABC$-model is chiral:
it contains the fields $A,B,C$, but not their conjugates $\bar A,\bar B,\bar C$.
Non-chiral instead are the kinetic terms $\Tr A\bar A, \ \Tr B\bar B$ and $\Tr C\bar C$.
The point is that non-chiral single-trace operators (\ref{complictedsingletrace})
can be obtained by the contraction of chiral and anti-chiral vertices with the
help of inverted kinetic terms (propagators), i.e. they can be produced
by tree Feynman diagrams. In terms of the Ward identities (or SDE), this is a corollary of the equations of motion
\be
\bar A+gBC=0,\ \ \ \ \ \ \ \bar B+gCA=0,\ \ \ \ \ \ \ \bar C+gAB=0
\ee
Consider just one example,
of how $\Tr (ABC)(\bar C\bar B\bar A)$ emerges from joining
two chiral and two antichiral vertices via three propagators:

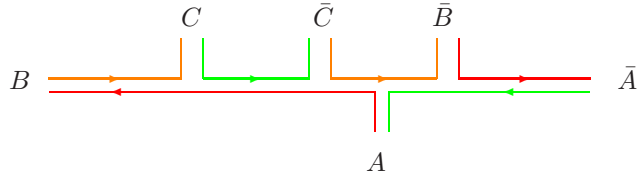
\begin{picture}(300,75)(-120,-40)
\put(-15,-4){\mbox{$B$}}
\put(50,20){\mbox{$C$}}
\put(100,20){\mbox{$\bar C$}}
\put(145,20){\mbox{$\bar B$}}
\put(215,-4){\mbox{$\bar A$}}
\put(120,-35){\mbox{$A$}}
{\co \put(0,0){\line(1,0){50}}\put(50,0){\line(0,1){15}}\put(25,0){\vector(1,0){2}}}
{\cg \put(55,0){\line(0,1){15}}\put(55,0){\line(1,0){40}}
\put(95,0){\line(0,1){15}}\put(75,0){\vector(1,0){2}}}
{\co \put(100,0){\line(0,1){15}}\put(100,0){\line(1,0){40}}
\put(140,0){\line(0,1){15}}\put(120,0){\vector(1,0){2}}}
{\cre \put(145,0){\line(1,0){50}}\put(145,0){\line(0,1){15}}\put(170,0){\vector(1,0){2}}}
\put(-13,0){
{\cre \put(0,-5){\line(1,0){123}}\put(123,-5){\line(0,-1){15}}\put(25,-5){\vector(-1,0){2}}}
{\cg \put(125,-5){\line(1,0){76}}\put(125,-5){\line(0,-1){15}}\put(170,-5){\vector(-1,0){2}}}
}
\end{picture}

\noindent
Chiral operators $\Tr (ABC)^k$ are not produced by a change of fields (integration
variables) from any other {\it chiral} operators, in this sense they are "cohomological".
Instead, they are variations of non-chiral kinetic terms,
\be
\Tr (ABC)^k = \delta (\Tr C\bar C) \ \ \ \ {\rm for} \ \ \ \
\delta \bar C = (ABC)^{k-1}(AB)
\ee
which have unit Jacobians.
Therefore, in the chiral sector, the Ward identities degenerate into
relations of the trivial type (\ref{derder})
and, in this sense, the dependence of effective action on the chiral couplings is
almost absent.
In fact, the chiral couplings
rather play the rode of the holomorphic {\it moduli} or "boundary conditions"
describing different solutions to the Virasoro constraints \cite{DV,AMM1,AMM2,CM,DVmore,M,MM}.
The corresponding integrability properties are not yet understood in universal terms.
While solutions of the Virasoro constraints are usually
integrable $\tau$-functions of non-chiral (non-cohomological) couplings \cite{MMint,UFN3,MMZ},
i.e. are fully controlled by Lie algebra representation theory,
dependence on the chiral (cohomological) couplings is rather similar to that on the
"point of the Grassmannian" \cite{MMint,UFN3,CM,RG,MM}.
It is presumably encoded in associated
quasiclassical (Whitham) $\tau$-functions and controlled by the WDVV equations \cite{WDVV,genWDVV,CM}.
Also, it is often described in a much poorer language of {\it chiral rings},
where chiral operators like $\Tr (ABC)^k$ form a multiplicative basis.
The theory of chiral effective actions is one of the directions
in which study of the rainbow tensor models
can produce considerable progress, simply because
{\bf the rainbow models are the first example among the topical matrix/tensor models which possesses
a non-trivial set of chiral operators.}

To avoid a terminological confusion: in a sense, chirality is no longer respected
in the higher rank tetrahedron/starfish models.
For example, for $D=3$ the fields can not be tensors of the type $(3,0)$ and $(0,3)$
(with all incoming or all outgoing arrows) only:
e.g., the tetrahedron vertex requires $(2,1)$ and $(1,2)$ fields
and has a shape $AB\bar C\bar D$ with two conjugate fields.
However,  if one calls chirality a counterpart of the above explained
"tree irreducibility" (which is the one relevant to the SUSY non-renormalization),
then, actually chiral are $A,B,\bar C,\bar D$, while $\bar A,\bar B, C, D$ are
anti-chiral fields,
and in this sense the tetrahedron/starfish interaction vertex is always {\it chiral}.

\paragraph{Ward identities.}
A counterpart of the Virasoro constraints in the tensor models was already
studied in \cite{GurVir} by a direct generalization of the standard line
of \cite{MMVir,UFN3}.
Needed for the derivation is a choice of shifts of integration variable (or considering a full derivative in the integrand) as
this defines a basis in the set of Ward identities.
A clever choice could be gradients of the chiral single trace operators.
Therefore, the Ward identities are closely related to the Connes-Kreimer Hopf algebra
\cite{CK}, which appears in the description of diffeomorphisms of the coupling constant
space in terms of Feynman graphs, in particular, in renormalization theory,
and is best understood and described in the basic matrix model terms \cite{GMS,MS}. Similarly, an example of tensor model was shown to be perturbatively renormalizable \cite{BR}, and the corresponding Connes-Kreimer Hopf algebra was constructed in \cite{RA} (see also some further development in \cite{ART}).

\paragraph{Spectral curve and AMM/EO topological recursion.}
The most popular choice of generating functions in matrix models is in terms
of resolvents \cite{KM,UFN3}
$$
\Tr\frac{1}{z-M} = \sum_{k=0}^\infty {1\over z^{k+1}}\,\Tr M^k
$$
In the large $N$ limit (or some subtler limits \cite{NS,MMMso}), the Ward identities can be often rewritten as a closed
algebraic (or differential/difference) equation for the average of the resolvent,
then, it is called spectral curve (or quantum spectral curve \cite{Ch,MMMso,GMM}) \cite{MM}, the equation looks
like (in the case of Hermitean one matrix model)
\be
W'(z) \left<\Tr \frac{1}{z-M}\right> + \left<\Tr\frac{1}{z-M}\right>^2 =
\sum_{a,b} T_{a+b+2}z^a \left<\Tr M^b\right> = f_W(z)
\ee
with $W(z) = \sum_k T_kz^k$ being the potential of matrix model.
The entire $1/N$ expansion then can be reformulated in terms of effective
field theory over the spectral curve and expressed through various quantities
on the spectral bundle over the moduli space. This procedure is known
as AMM/EO topological recursion \cite{AMMEO} (see also \cite{BD} for some tensor models).

However, the resolvent is not the only and often not the most clever choice:
for example, the use of the Wilson loops \cite{expgen}
$$\Tr e^{sM}=\sum_{k} \frac{s^k}{k!}\,\Tr M^k$$
and especially of the Harer-Zagier generating functions  \cite{HZ,MorSha,AMM1}
$$\sum_{N,k} \frac{\lambda^N z^k}{(2k-1)!!}\,{\rm Tr}_{N\times N} M^{2k}$$
leads to far more explicit expressions for matrix model averages.
Moreover, there is no direct generalization of resolvent from matrix to
tensor models, thus, one should consider generating functions in a somewhat more
abstract way.
In general, the Ward identities
are provided by a non-infinitesimal change of the integration variable,
which can be substituted by a certain shift of the coupling constants in
the partition function \cite{MMvir},
\be
Z\Big(T+v(T)\Big) =\  :e^{\hat v(T)}:\, Z(T),
\ee
with $\hat v(T) = v(T)\frac{\partial}{\partial T}$.
In every particular model, one can work out particular relevant diffeomorphisms
$v(T)$ which leave $Z$ intact,
but nice formulas usually require elimination of the normal ordering,
which is provided by a somewhat complicated
Bogolubov-Zimmermann  forest formula, see \cite{GMS} for details (see also \cite{Riv} for an interesting related issue).
Namely, without normal ordering the action of the vector field $\hat V$
involves a sum repeated actions of $V$ on itself:
\be
Z\Big(T+v(T)\Big)  = e^{\hat V} Z(T)
= \left(1+\sum_{{\rm forests}\ {\cal F}} \frac{1}{{\rm Tree}({\cal F})!}
\prod_{{\rm trees}\ {\cal T}\in{\cal F}} \frac{\hat V_{{\cal T}}}{\sigma_{{\cal T}}{\cal T}!}
\right)\, Z(T)
=\  :e^{\hat v(T)}:\, Z(T)
\ee
with certain $V$-independent combinatorial coefficients $\sigma_{\cal T}$ and ${\cal T}!$
This formula expresses $\hat v$ through $\hat V$.
The inverse transformation is best described, when the
partition function is expressed in terms of Feynman graphs,
$Z(T) = \sum_\Gamma F_\Gamma Z_\Gamma(T)$, then
 \be
e^{\hat V}\ F_\Gamma = \sum_{n=0}^\infty\left(
\sum_{\stackrel{\stackrel{{\rm non-intesecting}}{{\rm box-subgraphs}}
}{\gamma_1,\ldots,\gamma_n\in {\cal B}\Gamma}}
\hat v_{\gamma_1}\ldots\hat v_{\gamma_n}\ F_{\Gamma/(\gamma_1,\ldots,\gamma_n)}
\right)
\ee
where we refer to \cite{GMS} for definition of the box-subgraphs and other details.
What is important, this kind of expansions can be needed  for a generic
formulation of the Ward identities in tensor theories
and {\bf the Virasoro type constraints of \cite{Gurvir}
should be reformulated as acting on the Hopf algebra
of graphs}, which in our example are the Feynman diagrams in the rainbow model.

\paragraph{Conclusion.}

We made a brief review of the currently popular tensor models from the perspective of matrix model theory {\it a la} \cite{UFN3}.
Nowadays interest is concentrated on the tensor models of a very special
class with peculiar tetrahedron/starfish interactions,
where distinguished in the large $N$ limit are peculiar melonic diagrams,
while non-trivial planar diagrams are suppressed leading
to the drastic simplifications.
We have considered a universal model of this type, by making all possible
indices independent and extending the symmetry to its extreme, which
is a product of $\frac{D(D+1)}{2}$ unitary groups $U(N_a)$.
In this {\it rainbow} model, the non-melonic diagrams are not just
suppressed among the planar ones, but they simply do not contribute irrespective
of the value of $N$.
The large $N$ limit instead acquires another type of non-triviality:
a non-analytical dependence on ratios of different color numbers,
which is obscured and actually absent in rectangular matrix models and quiver
Yang-Mills theories, but which is present in non-perturbative description of
conformal theories.
The derivation of Ward identities is technically straightforward,
but, at a conceptual level, requires a suitable definition of generating
functions properly associated with the diffeomorphisms
group of the moduli space.
This puts the story within the context of the Bogoliubov-Zimmerman/Connes-Kreimer theory.
It is still a question whether a pronounced simplicity of the rainbow tensor models helps us to consider a new avenue of quantum gravity, but the study in this framework seems certainly important for development of group theory,
integrabilities and non-linear algebra as well.

\section*{Acknowledgements}

The work of H.I. was supported in part by JSPS KAKENHI Grant Number 15K05059. A.Mor. acknowledges the support of JSPS (\#  S16124) and hospitality of OCU. Our work was partly supported by RFBR grants 16-01-00291 (A.Mir.), 16-02-01021 (A.Mor.) and by joint grants
15-52-50041-YaF, 15-51-52031-NSC-a, 16-51-53034-GFEN, 16-51-45029-IND-a. Support from JSPS/RFBR
bilateral collaborations "Faces of matrix models in quantum field theory and statistical mechanics" is
gratefully appreciated.

\end{document}